\documentstyle[aps,prl,twocolumn,floats,epsfig]{revtex}
\begin{document}
\wideabs{
\title{Defect Turbulence in Inclined Layer Convection}
\author{Karen E. Daniels and Eberhard Bodenschatz\cite{eb:email} \\
Laboratory of Atomic and Solid State Physics, Cornell University,
Ithaca, NY 14853 }
\date{\today}
\maketitle

\begin{abstract}
We report experimental results on the defect turbulent state of
undulation chaos in inclined layer convection of a fluid with
Prandtl number $\approx 1$. By measuring defect density and
undulation wavenumber, we find that the onset of undulation chaos
coincides with the theoretically predicted onset for stable,
stationary undulations. At stronger driving, we observe a
competition between ordered undulations and undulation chaos,
suggesting bistability between a fixed-point attractor and
spatiotemporal chaos. In the defect turbulent regime, we measured
the defect creation, annihilation, entering, leaving, and rates. We
show that entering and leaving rates through boundaries must be
considered in order to describe the observed statistics. We derive 
a universal probability distribution
function which agrees with the experimental findings.
\end{abstract}
}

\pacs{
47.54.+r, 
47.20.-k, 
47.20.Bp, 
}

Weakly driven, dissipative pattern-forming systems often exhibit
the spatiotemporally chaotic state of defect turbulence, where the
dynamics of a pattern is dominated by the perpetual nucleation,
motion, and annihilation of point defects (or dislocations)
\cite{Coullet:1989:DMT}. Examples can be found within striped patterns
in wind driven sand, electroconvection in liquid crystals
\cite{Rehberg:1989:TWD}, nonlinear optics \cite{Ramazza:1992:STD},
fluid convection \cite{Morris:1993:SDC,LaPorta:2000:PMP},
autocatalytic chemical reactions \cite{Ouyang:1996:TFS}, and Langmuir 
circulation in the 
oceans \cite{Haeusser:1997:AMD}. The hope is that the dynamics of
these very different systems can be characterized by a universal
description which is based only on the defect dynamics. 

A first description of defect turbulence was given by Gil {\it et al.}
\cite{Gil:1990:SPD} for a spatiotemporally chaotic state of the
complex Ginzburg-Landau equation (CGLE).  They postulated that the
nucleation rate for defect pairs is independent of the number
of pairs $M$, and based on the topological nature of defects
the annihilation rate is proportional to $M^2$. Through detailed
balance, they showed that these assumptions lead to a 
squared Poisson distribution for the probability distribution
function (PDF) of $M$. They also found agreement with this PDF
in simulations of the CGLE with periodic  boundary conditions. 
Rehberg {\it et al.}  \cite{Rehberg:1989:TWD} measured the 
PDF of $M$ for defect 
turbulence in electroconvection of nematic liquid crystals and
found it to be consistent with the predicted squared Poisson
distribution. Later, Ramazza {\it et al.} investigated a
defect turbulent state in optical patterns and found that their
data was not conclusive. To date, studies in both simulation and
experiment have relied purely on comparisons of the PDFs. The
gain and loss rates, fundamental to the universal
description of defect turbulence, have not been measured. In
addition, effects due to boundaries were not considered.

In this Letter, we report experimental results on the
defect turbulent state of undulation chaos in inclined layer
convection of a fluid of Prandtl number  $\approx 1$. By
tracking all defects in a finite area of the convection cell we
measured, for the first time, defect creation, annihilation,
leaving, and entering rates for a defect turbulent state. The
observed pair creation and annihilation rates agree with the
predictions \cite{Gil:1990:SPD}. In an experimental system where
periodic boundary conditions do not apply, pairs of defects are
no longer an appropriate description. To describe the statistics
correctly, single defect leaving and entering rates through 
the boundaries must be considered. Our data show that
the entering rates are approximately 
independent of the number of positive/negative 
defects $N_\pm$ and the leaving rates are proportional to $N_\pm$. We
derive a universal PDF which reduces to the earlier predicted
squared Poisson distribution \cite{Gil:1990:SPD} when boundary
effects are negligible. Our measurements agree with this new
distribution. In addition, by measuring both $N_\pm$
and the undulation wavenumber we determined that the onset of
undulation chaos coincides with the theoretically predicted onset
for stable, stationary undulations. At higher driving, we observe
competition between regions of ordered undulations and
undulation chaos, suggesting bistability between a fixed point
attractor and  spatiotemporal chaos.

\begin{figure}[bt]
\centerline{\hfill (a) $\epsilon = 0.08$ \hfill \hfill 
(b) $\epsilon = 0.17$ \hfill \hfill (c) $\epsilon = 0.17$ \hfill}
\smallskip
\centerline{\epsfig{file={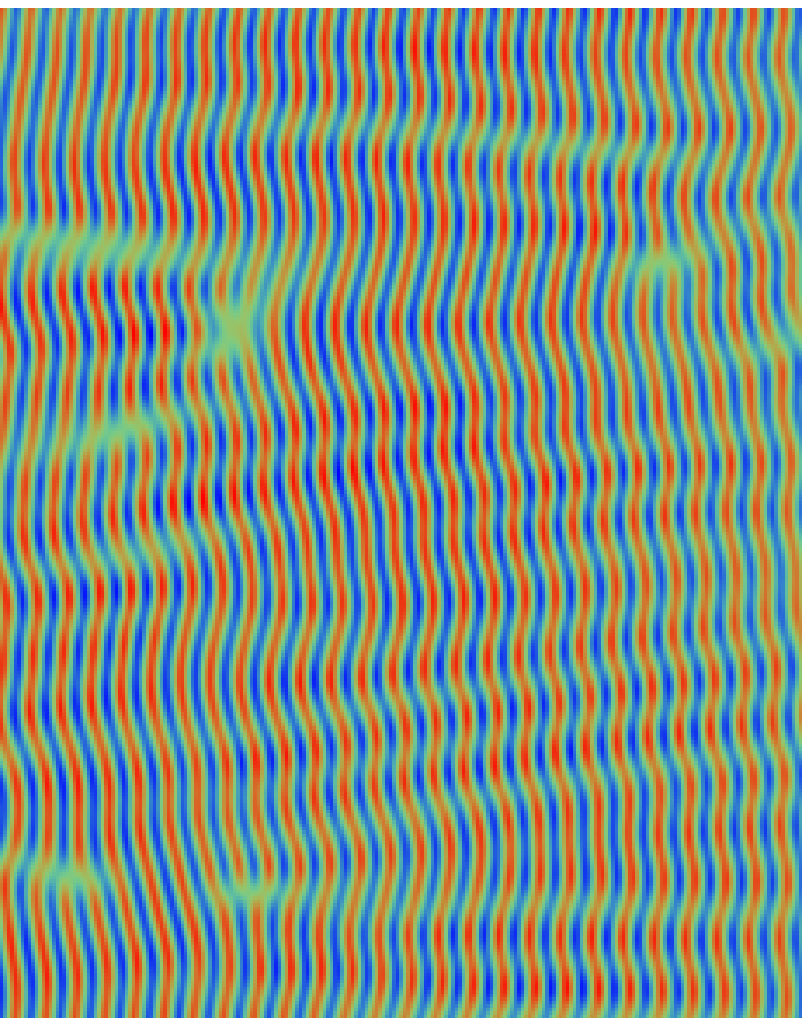}, width=1.1in}
    \epsfig{file={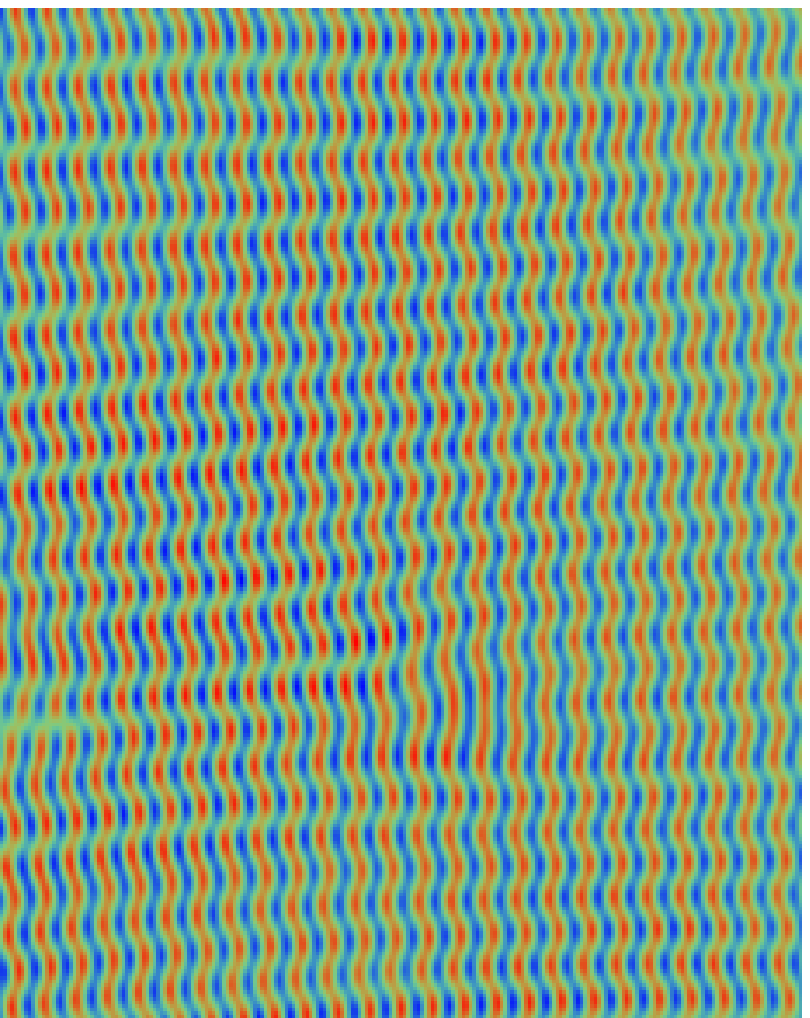}, width=1.1in}
    \epsfig{file={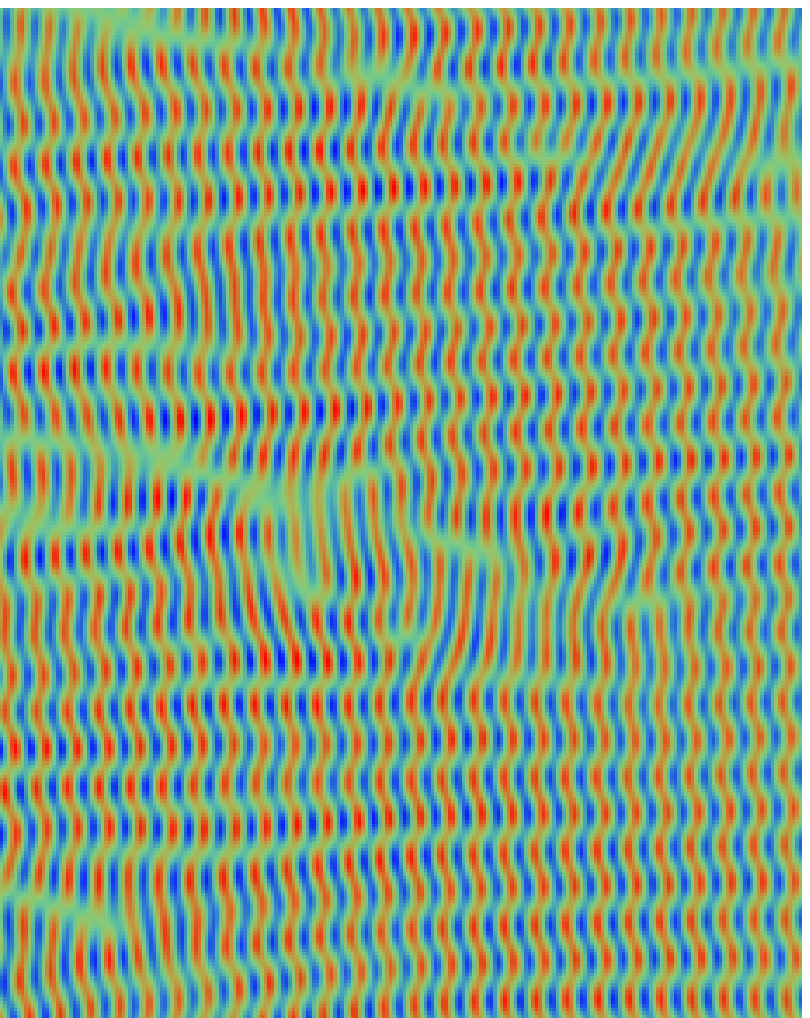}, width=1.1in}}
\caption{False color images of undulation chaos \protect\cite{EPAPS}
at $\theta = 30^\circ$. Red/blue are rising/falling
fluid and green is low-amplitude convection. Uphill 
direction is at the top of the page and the region shown is the
subregion of the cell used in the analysis. Positive defects point
uphill, negative defects downhill and the pattern drifts downward
\protect\cite{NOB} with a period $\approx 150 \tau_v$.}
\label{f_pic}
\end{figure}

When an inclined fluid layer is heated from below and cooled from
above for small temperature differences $\Delta T$ the fluid layer
experiences not only a linear
temperature gradient (as in Rayleigh-B\'enard convection)
but also a symmetry-breaking large scale
shear flow, with fluid rising at the warmer and falling
at the cooler plate \cite{Hart:1971:SFD}. Above a critical temperature
difference $\Delta T_c$, a
pattern of longitudinal rolls aligned with the uphill/downhill
direction is observed for intermediate inclination angles. 
For $\epsilon \gtrsim 0.02$, where
$\epsilon \equiv \Delta T / \Delta T_c - 1$, these longitudinal
rolls are unstable to undulations 
\cite{Hart:1971:SFD,Clever:1977:ILC}. While it was
predicted theoretically that a perfect pattern of undulations is
stable \cite{Pesch+Brausch}, a defect turbulent state of
undulation chaos was found experimentally 
\cite{Daniels:2000:PFI}. Recently, undulation
chaos was also observed in numerical simulations of the Navier-Stokes
equations when random initial conditions were used
\cite{Pesch+Brausch}, while with controlled initial conditions undulations
were found to be
stable. As shown in the snapshot in Fig.~\ref{f_pic}a, undulation
chaos is characterized by a pattern of disordered undulating rolls,
punctuated by point defects which carry a topological
charge of $\pm 2\pi$ \cite{Cross:1993:PFE}. Defects are nucleated
pairwise in regions of low convective amplitude (green), 
enter and leave the system at the boundaries, preferentially move at
right angles to the rolls, and annihilate as pairs.  For $\epsilon
\gtrsim 0.10$, the defect turbulent state appears intermittently with
ordered regions of undulations, suggesting a competition between the
fixed point attractor (undulations) and spatiotemporal chaos
(undulation chaos). This is shown by the two snapshots in
Fig.~\ref{f_pic}b, c. The dynamics of the state is best viewed in the
movies available at Ref.~\cite{EPAPS}.

The experiments were conducted in the same apparatus as 
Ref.~\cite{Daniels:2000:PFI} for an inclination angle of $\theta =
30^\circ$. The fluid was compressed CO$_2$ at a pressure of $(56.5
\pm 0.01)$ bar regulated to $\pm 0.005$ bar with a mean
temperature of $(28 \pm 0.05) ^\circ$C regulated to $\pm
0.0003^\circ$C. For these parameters, convection appeared at
$\Delta T_{\mathrm c} = (1.763 \pm 0.005) ^\circ$C. The planform of the
convection pattern was observed by the usual shadowgraph technique
\cite{deBruyn:1996:ASR}.  
A cell of  height $d = (388 \pm 2) \mu$m and
dimensions $101d \times 50d$ was used, for which
the vertical diffusion time was $\tau_v
\approx 1.3$ sec. All data was measured in a
homogeneous central subregion of size $32d \times 25d$
\cite{subregion_note}. To track defects, runs of 100 shadowgraph images
were taken at a rate of 3 frames/sec ($\approx \tau_v / 4$) with a
10 bit 1K $\times$ 1K digital CCD camera. For 17 values of
$\epsilon$ between 0.04 and 0.22 this was repeated for 600 (500, 400)
runs at $\epsilon \le 0.07$ 
($0.08 \le \epsilon \le 0.10$, $\epsilon \ge 0.12$). 
Runs at the same $\epsilon$ were separated by at least $100 \tau_v$
for statistical independence. A total of
1.5 terabytes of data were analyzed. Each value of $\epsilon$ was
reached via a quasistatic temperature increase in steps of $0.001
^\circ$C, waiting $700 \tau_v$ between steps 
for transients to die out. We also conducted a sequence of
measurements with quasistatic temperature decreases to check for
possible hysteresis, which was not observed.

To lowest order, the planform of undulation chaos can be captured by
the real part of the two dimensional field $\Psi(x,y)$ with
\begin{equation}
\Psi = \psi_0 e^{i (q x + \phi)} +
i \psi_1 \left ( e^{i(qx+ py)} + e^{i(qx-py)} \right ) + \cdots
\label{e_perfect}
\end{equation}
where the wavenumbers $q$ and $p$ describe the stripe spacing and
waviness, respectively. Both the wavenumbers and the
amplitudes $\psi_0$ and $\psi_1$ vary slowly
in space and time. For a perfect undulation pattern $\phi$ is constant
everywhere, while for undulation chaos $\phi(x,y,t)$ approximates
the deformation of the pattern due to the defects. The ansatz
$\phi(x,y,t) = \sum c_i \arctan({y - y_i \over x - x_i})$ describes
the observed pattern well for constant $q, p,$ and $ \psi_1 / \psi_0$,
where $(x_i,y_i)$ are the locations
of  defects and $c_i = \pm 1$ are their  corresponding topological
charges.

\begin{figure}
\centerline{ \epsfig{file={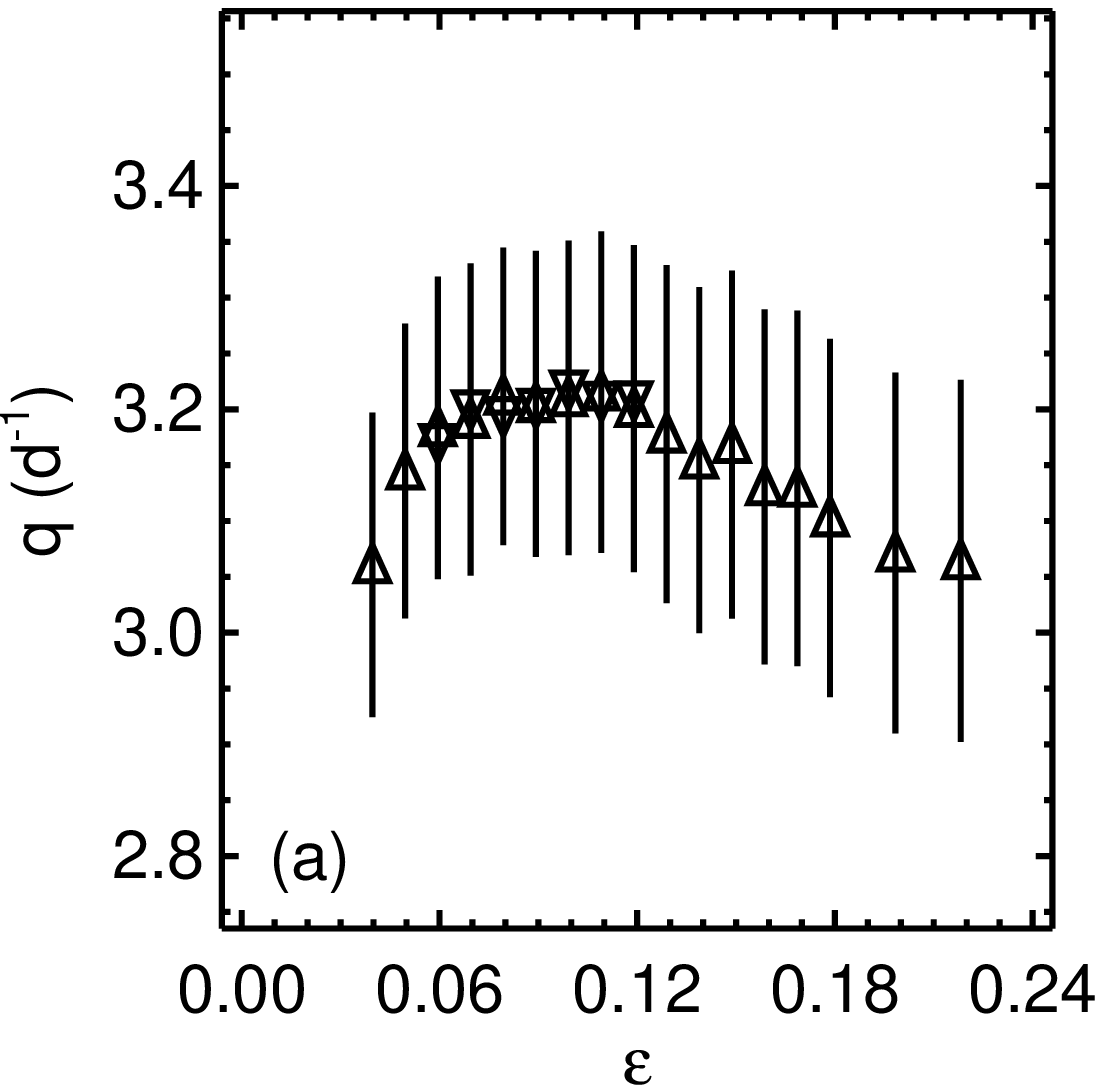}, height=1.5in}~~
\epsfig{file={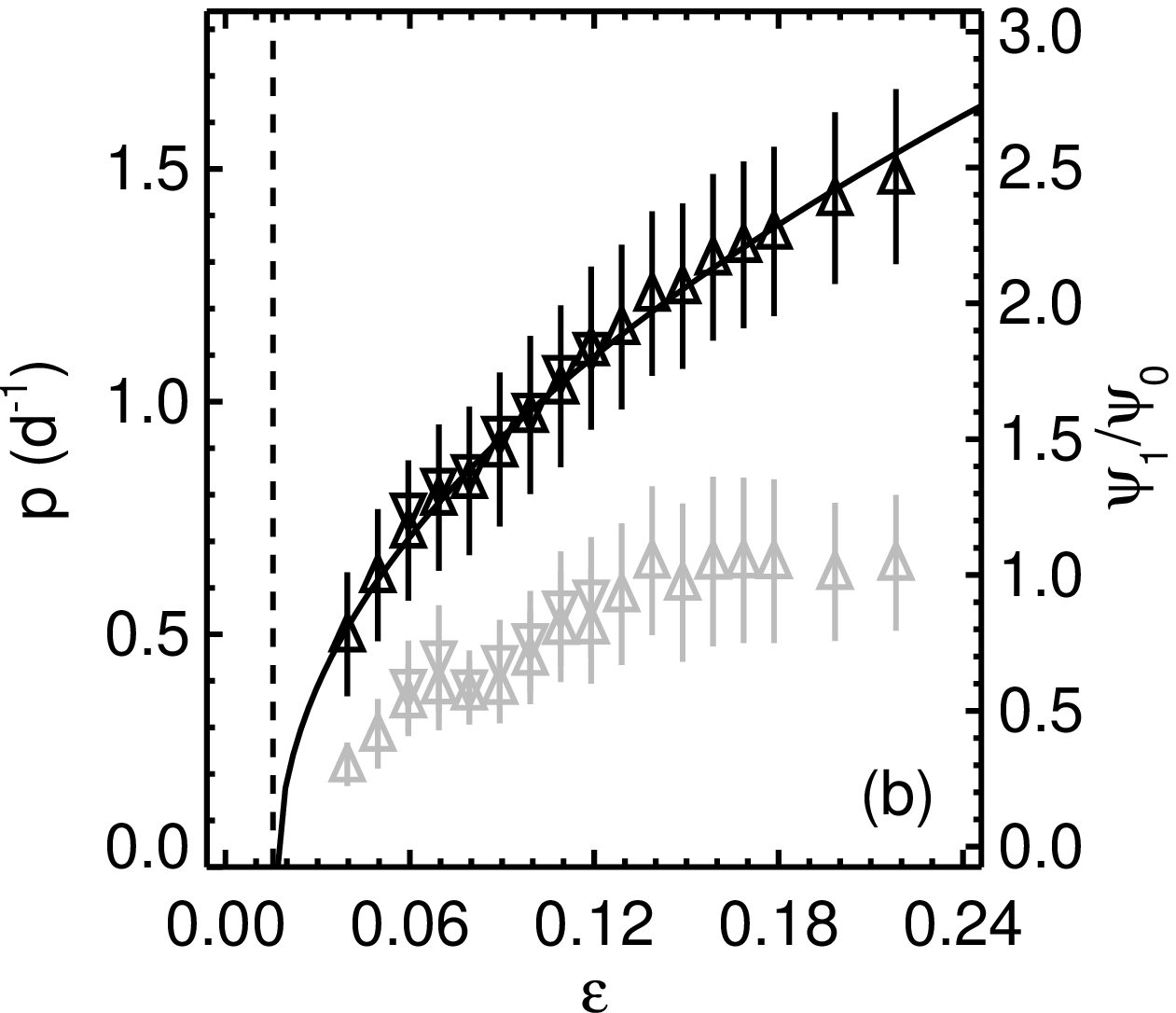}, height=1.5in}} 
\caption{(a) Roll wavenumber $q$ and (b) undulation wavenumber $p$
(black) with relative amplitude  $\psi_1/\psi_0$ (gray), as a function
of $\epsilon$. Values are taken at the peaks of the PDF, with bars
representing the width of the distribution at one standard deviation. 
Data taken while increasing temperature are shown with symbol
$\bigtriangleup$ and decreasing with $\bigtriangledown$. Dashed
line represents prediction for onset from \protect\cite{Pesch+Brausch}.}
\label{f_dist}
\end{figure}

In Fig.~\ref{f_dist} the peak values for the distributions of $p$,
$q$, and $\psi_1/\psi_0$ are shown as a function of $\epsilon$
\cite{localwn}.
The values for $q$ change little with $\epsilon$, while the undulation
wavenumbers can be fit by $p \propto \sqrt{\epsilon -
\epsilon_u}$, shown as a solid black line. This fit provides a
value for the onset of undulation chaos at $\epsilon_u = 0.017 \pm
0.001$, which is in good agreement with $\epsilon_u = 0.016$
predicted theoretically for the onset of stable stationary
undulations \cite{Pesch+Brausch}.

\begin{figure}[tb]
\centerline{\epsfig{file={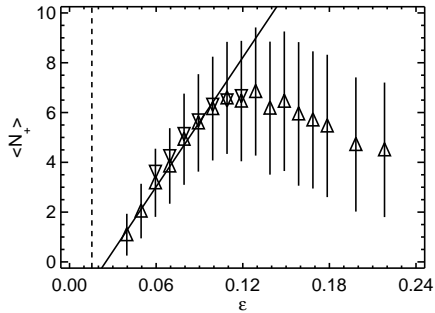}, width=2.3in}}
\caption{Number of positive defects in the subregion as a function of
$\epsilon$, with bars representing the width of the distribution
at one standard deviation.} \label{f_denseps}
\end{figure}

In order to better characterize the generic properties of
defect turbulence, we determined defect locations and the
corresponding topological charges in each shadowgraph image. A
time trace of the number of defects
over $ 2 \times 10^5 \tau_v$ at $\epsilon = 0.13$ ($10^4$ defects)
revealed that the undulation chaos state is statistically
stationary and neutral with respect to topological
charge. Nonetheless, any single realization of the 
pattern does not necessarily satisfy $N_+ = N_-$ due to the
boundaries. From the list of
defect locations in each frame, we determined the mean $\langle N_\pm
\rangle$ and the variance   
$\sigma_\pm^2 = \langle N_\pm^2 \rangle - \langle N_\pm \rangle^2$  
in the chosen subregion, plotted in Fig.~\ref{f_denseps}
as a function of $\epsilon$. Here and below, we report results only
for $N_+$; equivalent results were obtained for $N_-$.  

For $\epsilon \le 0.10$, $N_\pm(\epsilon)$  is linearly increasing.
Extrapolation to $N_+ = 0$ gives $\epsilon_d = 0.025 \pm 0.002$,
which is close to the onset of undulations ($\epsilon_u = 0.017
\pm 0.001$) determined above from wavenumber measurements. 
The difference may be attributable to the finite size of the
system. For $\epsilon > 0.10$, $N_\pm(\epsilon)$ decreases in the
bistability regime, while no such decrease is seen in $p$.
 
Two predicted generic features of defect turbulence
\cite{Gil:1990:SPD} are the constant creation rate and the
annihilation rate quadratic 
in the number of defect pairs $M$. By connecting defect locations in
adjacent frames into defect tracks, and then associating tracks of
opposite sign to locate creation and annihilation events, we were able
to perform the first direct test of the theoretical postulates, as
well as the entering and leaving rates to/from the observation area. 
For a system where single defects can enter and leave through the
boundaries a pair-based  description is no longer justified. Thus, we
separately consider the PDFs for $N_\pm$.

\begin{figure}[bt]
\centerline{\epsfig{file={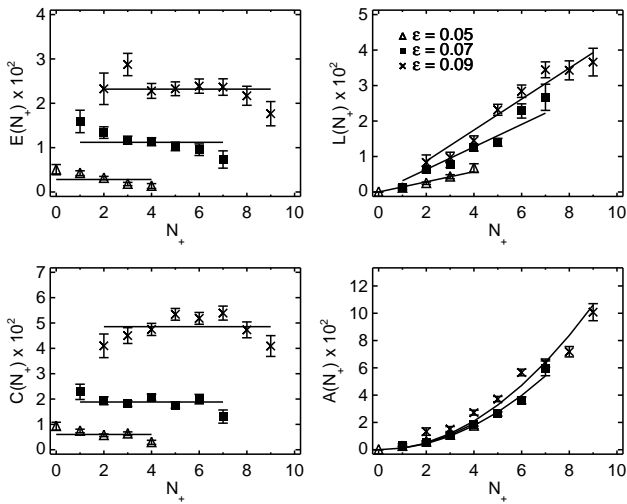}, width=3.375in}}
\caption{Probability of entering, and leaving, creation, and
annihilation (per 0.33 sec)
as a function of $N_+$ for several $\epsilon$. Lines are fits to
Eqn.~\protect\ref{e_rates}.}
\label{f_4rates}
\end{figure}

The creation rate $C$ was observed to be 
approximately independent of $N_\pm$ 
and the annihilation rate agreed with $A \propto N_+ N_-$, by
extension from $A \propto M^2$ for the topologically neutral case.
Fig.~\ref{f_4rates} shows the four gain/loss rates, tabulated using
positive defects and events only. To lowest order, the entering rate 
is independent of the number of defects already in the subregion
of the cell, suggesting that both creation and entering are random
processes. The leaving rate was found to increase in proportion to
$N_\pm$ as expected, since unlike annihilations such events do not
depend on the presence of two oppositely-charged defects.

The observed rates are  approximately given by
\begin{eqnarray}
\nonumber E(N_\pm) &=& E_0 \\
\nonumber C(N_\pm) &=& C_0 \\ 
          L(N_\pm) &=& L_0 N_\pm \label{e_rates} \\
\nonumber A(N_\pm) &=& A_0 N_\pm^2
\end{eqnarray}
which consider defects of a single sign only.
The fits corresponding to Eqn.~\ref{e_rates} are shown in 
Fig.~\ref{f_4rates}, for three values of $\epsilon$. It is important to
note that this is a simplification. $C(N_\pm)$
and $E(N_\pm)$ show a weak, negative slope: defects are slightly less
likely to enter or be created if there are already defects
present. We also observe a weak dependence of these rates on the
number of oppositely-charged defects: $E(N_+)$ is slightly 
diminished and $L(N_+)$ is slightly elevated for $N_+ > N_-$. Thus, 
to fully describe the system, we would need to consider rates 
and distributions for both
$N_+$ and $N_-$ simultaneously. In order to make analytical progress we 
will assume below that Eqn.~\ref{e_rates} are a complete
description. As will be shown below, these assumptions approximate the
data well.

Using Eqn.~\ref{e_rates}, we can construct the PDF
for $N \equiv N_\pm$. Assuming a stationary distribution, detailed
balance requires ${\mathrm loss}(N) {\cal P}(N) = {\mathrm gain}(N-1)
{\cal P}(N-1)$ to describe the probabilities at adjacent $N$
\cite{time_reverse}.  For the relevant rates,
\begin{eqnarray}
{\cal P}(N) & = & { E(N) +C(N) \over L(N) + A(N) } {\cal P}(N-1)
\label{e_modgen}\\
\nonumber   & = & {\alpha \over \beta N + N^2} {\cal P}(N-1).
\end{eqnarray}
where $\alpha \equiv {E_0 + C_0 \over A_0}$ and $\beta \equiv {L_0
\over A_0}$. By performing the recursion and properly normalizing
the distribution, we find a {\it modified} Poisson distribution
\begin{equation}
{\cal P_{\alpha, \beta}}(N) = {\alpha^{{\beta \over 2} + N} \over
I_\beta(2 \sqrt{\alpha}) \Gamma(1 + \beta +N) N!}
\label{e_modpoiss}
\end{equation}
where $I_\beta$ is the modified Bessel function. For $\beta= 0$ 
($\alpha  = \langle N^2 \rangle$) this PDF reduces to
the squared Poisson distribution of Ref.~\cite{Gil:1990:SPD}. 
In our experiments, $\beta \approx 3$ \cite{Gaussian}.

\begin{figure}
\centerline{\epsfig{file={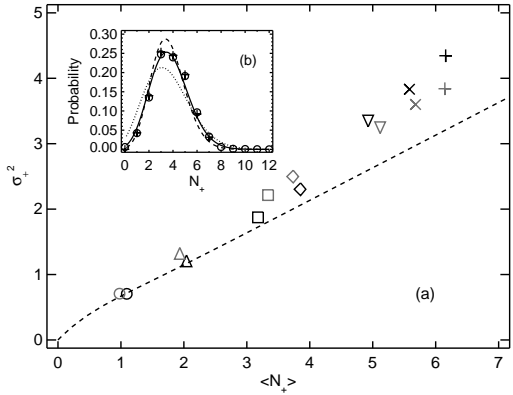}, width=2.7in}}
\caption{ (a) Mean number of positive defects
{\it vs.} $\sigma_+^2$ for $\epsilon$ between 0.04 and 0.10.
Black symbols are experimental; gray symbols are from 
Eqns.~\protect\ref{e_mean} and \protect\ref{e_second}; dashed line
is the analytical relation for squared Poisson distribution 
\protect\cite{Gil:1990:SPD}.
(b) PDF for $6 \times 10^4$ defects at $\epsilon=0.07$, with four
distributions shown for comparison. $\bigcirc$ are the
experimental distribution, solid line is modified Poisson,
dashed is squared Poisson, dotted is Poisson, and $+$ are from 
the raw rates in Fig.~\protect\ref{f_4rates}.}
\label{f_poiss}
\end{figure}

Fig.~\ref{f_poiss}b shows that the squared Poisson and Poisson
distributions match the data poorly. Both the modified Poisson
distribution and a distribution obtained by performing the
recursion of Eqn.~\ref{e_modgen} on the raw experimental data from
Fig.~\ref{f_4rates} agree well with the experimental data.
The modified Poisson distribution also correctly captures the mean
and width of the PDF. Using the already normalized
 modified Poisson distribution 
(Eqn.~\ref{e_modpoiss}), we calculate $\langle N \rangle = \sum{N {\cal
P}_{\alpha,\beta}(N)} $ and $\langle N^2 \rangle = \sum{N^2 {\cal
P}_{\alpha,\beta}(N)}$ and obtain the analytic expressions
\begin{eqnarray}
\langle N \rangle & = & \sqrt{\alpha} { I_{\beta+1}(2 \sqrt{\alpha})
\over I_\beta(2 \sqrt{\alpha}) } \label{e_mean} \\
\langle N^2 \rangle & = &
\sqrt{\alpha} { I_{\beta+1}(2 \sqrt{\alpha}) +
\sqrt{\alpha} I_{\beta+2}(2 \sqrt{\alpha})
\over I_\beta(2 \sqrt{\alpha}) } \label{e_second}.
\end{eqnarray}
Again, the above
equations reduce to the corresponding equations for the squared
Poisson distribution in the limit $\beta = 0$.
In Fig.~\ref{f_poiss}a the dependence of the mean on the variance
is plotted for the raw data and the modified
and squared Poisson distributions.  The modified
Poisson distribution (Eqn.~\ref{f_poiss}) shows excellent
agreement, while the squared Poisson distribution clearly fails to
describe the experimental data.

In summary, we find that the state of undulation chaos, and not
the fixed point attractor to ordered undulations, is selected
above the predicted onset for undulations. At higher driving,
we find an apparent competition between the two attractors. 

Our result for the $N$-dependence of the defect creation,
annihilation, leaving, and entering rates and the modified Poisson
(Eqn.~\ref{e_modpoiss}) should be generic to any
defect turbulent system. We have found agreement with the
theoretical predictions for creation and annihilation
rates \cite{Gil:1990:SPD} and have extended 
the analysis to the experimentally relevant case of a finite system
without periodic boundary conditions where it is important to consider
both positive and negative defects (rather than pairs). Although we
observed weak fluctuations away from topological charge neutrality, a
description  based purely on a single topological 
charge was sufficient to describe the observed behavior \cite{inprep}.

We thank W. Pesch and O. Brausch for sharing the results of
simulations and stability analysis which have enhanced our
understanding of the bistability. We would like to thank the NSF for
support under DMR-0072077.

\bibliographystyle{prsty}

\begin{thebibliography}{10}
\bibitem{subregion_note}
This was the largest region which showed random defect tracks. For smaller
  observation areas, the statistics behaved equivalently.
\bibitem{time_reverse}
Analysis of time-reversed raw data reveals that the time symmetry required for
  detailed balance is satisfied.
\bibitem{Gaussian}
For large $N_\pm$, the PDF becomes Gaussian and the underlying mechanisms are
  obscured.
\end{thebibliography}

\end{document}